\documentclass[11pt]{article}
\usepackage[final]{acl}
\usepackage{times}
\usepackage{latexsym}
\usepackage[T1]{fontenc}
\usepackage[utf8]{inputenc}
\usepackage{microtype}
\usepackage{inconsolata}
\usepackage{graphicx}
\usepackage{booktabs}
\usepackage{amsmath}
\usepackage{amssymb}
\usepackage{url}
\usepackage[section]{placeins}
\title{RouteGuard: Internal-Signal Detection of Skill Poisoning in LLM Agents}

\author{
\begin{tabular}{c}
{\normalfont\mdseries\small Wenjie Xiao$^{1,2}$ \hspace{0.9em} Xuehai Tang$^{2,*}$ \hspace{0.9em} Biyu Zhou$^{2}$ \hspace{0.9em} Songlin Hu$^{1,2}$ \hspace{0.9em} Jizhong Han$^{1,2}$} \\
\\[0.5em]
{\normalfont\mdseries $^{1}$ University of Chinese Academy of Sciences} \\
{\normalfont\mdseries $^{2}$ Institute of Information Engineering, Chinese Academy of Sciences} \\
{\normalfont\mdseries\texttt{xiaowenjie@iie.ac.cn, tangxuehai@iie.ac.cn, zhoubiyu@iie.ac.cn,}} \\
{\normalfont\mdseries\texttt{husonglin@iie.ac.cn, hanjizhong@iie.ac.cn}} \\
{\normalfont\mdseries $^{*}$ Corresponding author}
\end{tabular}
}

\begin{document}
\maketitle
\begin{abstract}
Agent skills are rapidly becoming the reusable interface layer of LLM agents, but they also create a new supply-chain-style indirect injection threat: an attacker can hide malicious instructions inside a skill that already looks like legitimate guidance and thereby hijack the agent's goal. We study the pre-execution detection problem: \emph{given an untrusted skill, how can we detect, inside an instruction-like skill carrier, malicious instructions that can redirect model reasoning and induce erroneous behavior before the skill is executed?} This setting is fundamentally different from ordinary indirect prompt injection, because benign skills are already dense, action-oriented instruction sources. As a result, rule-based matching and surface semantic screening become brittle exactly in the slices that matter most.

We show that successful skill poisoning produces a structured internal effect that we call \textbf{attention hijacking}: during response generation, attention shifts from trusted context to the malicious skill span, the shift is stronger for earlier injections and for \texttt{description}-channel payloads, and larger shifts are associated with harmful outputs. Motivated by this mechanism, we propose \texttt{RouteGuard}, a frozen-backbone detector that combines response-conditioned attention and hidden-state alignment signals through reliability-gated late fusion. Across five evaluation questions spanning heterogeneous skill benchmarks and ordinary indirect prompt injection, \texttt{RouteGuard} is consistently the strongest or most robust detector; on the critical \texttt{Skill-Inject} channel slice, it reaches \texttt{0.8834} F1 and recovers \texttt{90.51\%} of \texttt{description} attacks missed by lexical screening. The central conclusion is simple: skill poisoning should be treated as malicious instruction competition inside an instruction-like carrier, and reliable defense therefore requires internal-signal detection rather than text-only filtering.
\end{abstract}

\section{Introduction}
Agent skills, marketplace plugins, and local capability bundles are rapidly becoming the reusable interface layer of LLM agents. This shift improves capability composition, but it also creates a new supply-chain-style indirect injection threat: once a malicious skill is installed, the same poisoned artifact can be invoked across many downstream tasks and can \textbf{hijack the agent's goal} long after the original installation event. The security problem is therefore not a one-shot prompt perturbation. It is pre-execution screening of reusable control artifacts.

The central difficulty is structural, not merely lexical. Benign skills are already \textbf{instruction-like carriers}: they tell the agent what to do, when to invoke the skill, and which constraints to follow. Unlike ordinary indirect prompt injection, where malicious text is often inserted into evidence-like content such as web pages or retrieved passages, skill poisoning blends malicious instructions into text that is already semantically licensed to guide behavior. The boundary between benign and malicious spans is therefore weak exactly where the attack matters most.

Existing defenses largely follow three paradigms. \textbf{Rule-based scanners} look for explicit signatures such as suspicious strings, dangerous API patterns, or handcrafted heuristics. \textbf{Semantic detectors} ask whether a passage ``looks like'' malicious instruction. \textbf{Hybrid approaches} combine the two. These methods can be useful when the attack appears as a clear anomaly inside neutral evidence. They become brittle when the carrier itself is instruction-bearing, because benign and malicious spans then become lexically and semantically homogeneous in the very region where the attack hides. Recent attention-based defenses further show that internal attention can be useful for \emph{generic} indirect prompt-injection detection and sanitization in untrusted external data \citep{zhong2025attentionall}. However, they mainly establish that attention is useful, without deeply analyzing why it is useful, and their setting does not extend cleanly to instruction-like skill carriers. Our paper is centered on that gap.

\paragraph{Research Question.}
This paper asks a sharper question than generic prompt-injection detection: \emph{given an untrusted skill, how can we detect, inside an instruction-like skill carrier, malicious instructions that can redirect model reasoning and induce erroneous behavior before the skill is executed?} The phrase ``instruction-like carrier'' is not rhetorical. It names the central difficulty of the problem. A poisoned skill is dangerous precisely because the carrier already looks like a legitimate instruction source.

Our answer starts from mechanism rather than classifier engineering. If a poisoned skill succeeds, it should not succeed merely because it contains suspicious tokens. It should succeed because it changes the model's internal control allocation during generation. We call this effect \textbf{attention hijacking}: response-time attention shifts away from trusted task-relevant context and toward a small untrusted skill region. Controlled analysis shows that this shift is structured rather than random, larger for earlier injections, much larger for \texttt{description}-channel poisoning than for direct line edits, and positively associated with harmful output behavior.

Only after establishing that motivation chain do we derive the detector. \texttt{RouteGuard} keeps the backbone frozen, probes the model with purpose-conditioned prompts, extracts response-conditioned attention and hidden-state alignment features, and combines them through reliability-gated late fusion. The design goal is not to overclaim a universal attention-only defense. It is to build a detector whose features and decision rule follow directly from the failure mode that the motivation study reveals. Across heterogeneous skill benchmarks, \texttt{RouteGuard} is the only method that remains strong everywhere; on the most critical \texttt{Skill-Inject} channel slice, it reaches \texttt{0.8834} F1 and recovers \texttt{90.51\%} of \texttt{description} attacks that lexical screening misses.

This paper makes four contributions:
\begin{enumerate}
    \item \textbf{Problem formulation.} We formulate skill poisoning as malicious-instruction detection inside an instruction-like carrier, clarifying why the problem is structurally different from ordinary indirect prompt injection and why the practical risk is \textbf{goal hijacking} rather than generic prompt corruption.
    \item \textbf{Mechanistic evidence.} We present a motivation study showing that successful skill poisoning manifests as structured attention hijacking whose magnitude depends on injection position and semantic channel and correlates with harmful behavior.
    \item \textbf{Detector design.} We propose \texttt{RouteGuard}, a frozen-backbone detector that combines response-conditioned attention and hidden-state alignment through hierarchical chunking, multi-probe observation, and reliability-gated late fusion.
    \item \textbf{Evaluation.} We evaluate \texttt{RouteGuard} through a question-driven protocol covering cross-benchmark comparison, slice analysis, expert ablation, comparison with traditional indirect prompt-injection detectors, and transfer back to ordinary indirect prompt injection.
\end{enumerate}

\paragraph{Problem Setup.}
Let $x^{\mathrm{tr}} = (x_{\mathrm{sys}}, x_{\mathrm{usr}})$ denote the trusted system and user instructions, let $s$ denote an untrusted skill artifact, and let $y \in \{0,1\}$ indicate whether $s$ contains malicious instructions that can redirect model reasoning and induce erroneous behavior. The task is \textbf{pre-execution skill-poison detection}: predict $y$ before the skill is executed. This formulation is intentionally narrower than marketplace-wide repository auditing or full runtime prevention. Its distinctive difficulty is that $s$ is an \emph{instruction-like carrier}: even benign skills contain action-oriented language, so poisoned skills do not have to appear as clean lexical anomalies inside a neutral context.

Given a frozen backbone $\theta$ and a short fixed probe continuation $x_{\mathrm{resp}}$, one probe forward pass exposes internal features
\[
z = \Psi_{\theta}(x^{\mathrm{tr}}, s, x_{\mathrm{resp}}) \in \mathbb{R}^{d},
\]
where $\Psi_{\theta}$ may include response-conditioned attention, head-level routing, and hidden-state statistics. A detector $g_{\phi}(z) \in [0,1]$ outputs a risk score, and the deployment decision is
\[
\hat{y} = \mathbf{1}[g_{\phi}(z) \ge \tau].
\]
The detector is evaluated under a deployment-aware objective:
\[
\begin{aligned}
\text{minimize}\quad & \Pr(\hat{y}=0 \mid y=1) \\
\text{subject to}\quad & \Pr(\hat{y}=1 \mid y=0) \le \alpha,
\end{aligned}
\]
that is, minimize attack miss subject to a benign-block budget $\alpha$. The scientific question is therefore whether frozen-model internal signals can support more reliable detection than rule matching and surface semantic screening when malicious instructions are hidden inside an already instruction-like skill carrier.

\section{Motivating Analysis}
The motivation follows a three-layer argument. First, skill poisoning is not ordinary indirect prompt injection, because the carrier itself is already instruction-like. Second, when a malicious instruction competes successfully inside such a carrier, it induces an internal routing change that can \textbf{hijack the agent's goal}. Third, because this shift is behaviorally relevant and surface boundaries are weak, a practical detector should use internal signals and fuse complementary experts rather than rely on text-only screening.

\subsection{Instruction-Like Carriers, Not Ordinary Indirect Prompt Injection}
\begin{figure}[t]
    \centering
    \includegraphics[width=1.04\columnwidth]{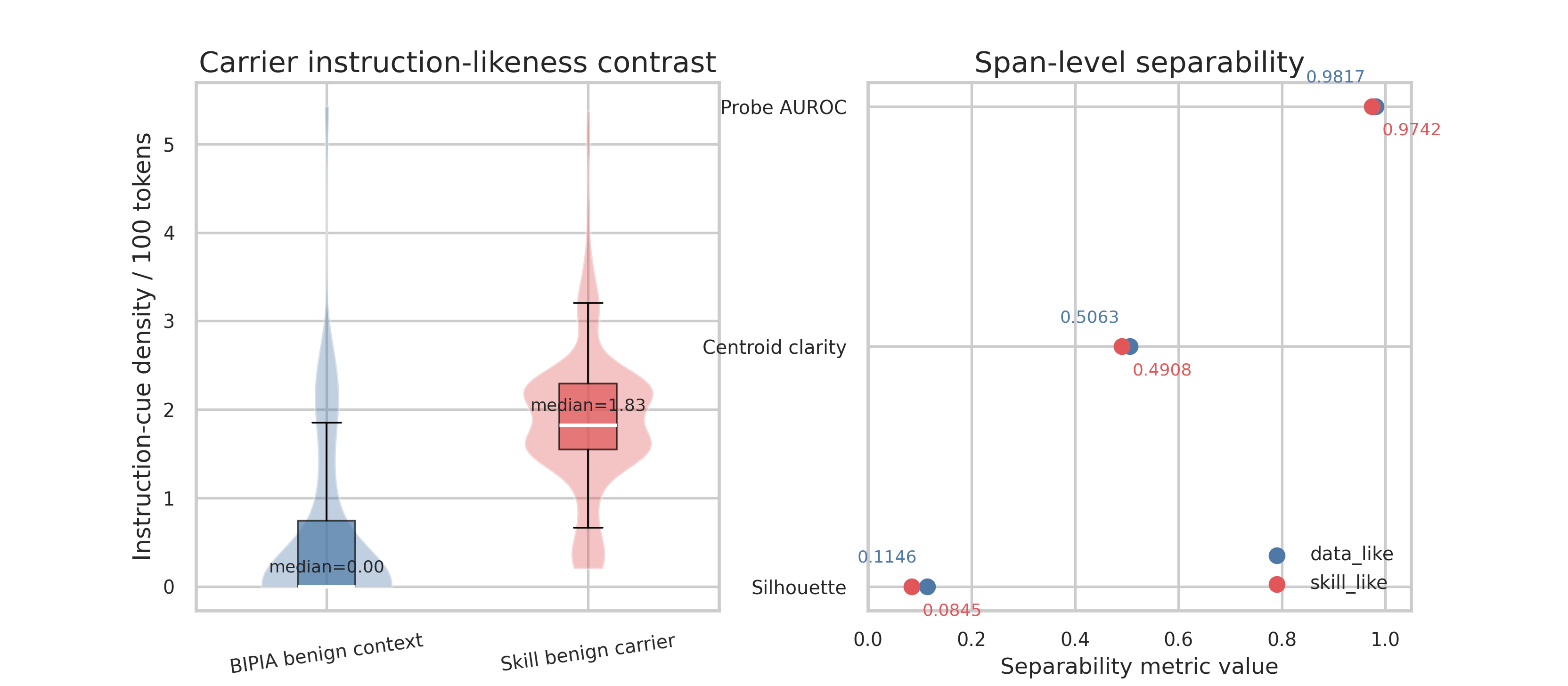}
    \caption{Boundary contrast between skill poisoning and ordinary indirect prompt injection. Benign skill carriers are already more instruction-like than benign external contexts, lexical separation is weaker in the subtle contextual slice, and white-box contrasts show more blended competition rather than a clean anomaly boundary.}
    \label{fig:motivation_boundary}
\end{figure}

The first question is whether skill poisoning should simply be folded into generic indirect prompt injection. Figure~\ref{fig:motivation_boundary} shows why the answer is no: the benign carrier in skill poisoning already looks like an instruction source, which makes malicious content harder to isolate both on the surface and in representation space.

The conclusion is straightforward. A skill is not ordinary evidence-like content; it is already semantically licensed to instruct behavior. This weakens the boundary between benign and malicious spans before any attack-specific reasoning even begins. As a result, the decisive failure mode is not ``malicious text appears in context,'' but ``malicious instructions become competitive inside a trusted-looking instruction carrier.''

This distinction is what separates skill poisoning from ordinary IPI. Ordinary IPI is closer to anomaly insertion into evidence-like content. Skill poisoning is closer to malicious instruction competition inside a legitimate instruction-bearing carrier. That difference is scientifically more important than the slogan ``harder,'' and it directly explains why the contextual skill-poisoning slice is the one where text-only detection degrades most sharply. The detailed lexical and white-box statistics are deferred to Appendix~A.

\subsection{Attention Hijacking Leads to Goal Hijacking}
\begin{figure*}[t]
    \centering
    \includegraphics[width=0.60\textwidth]{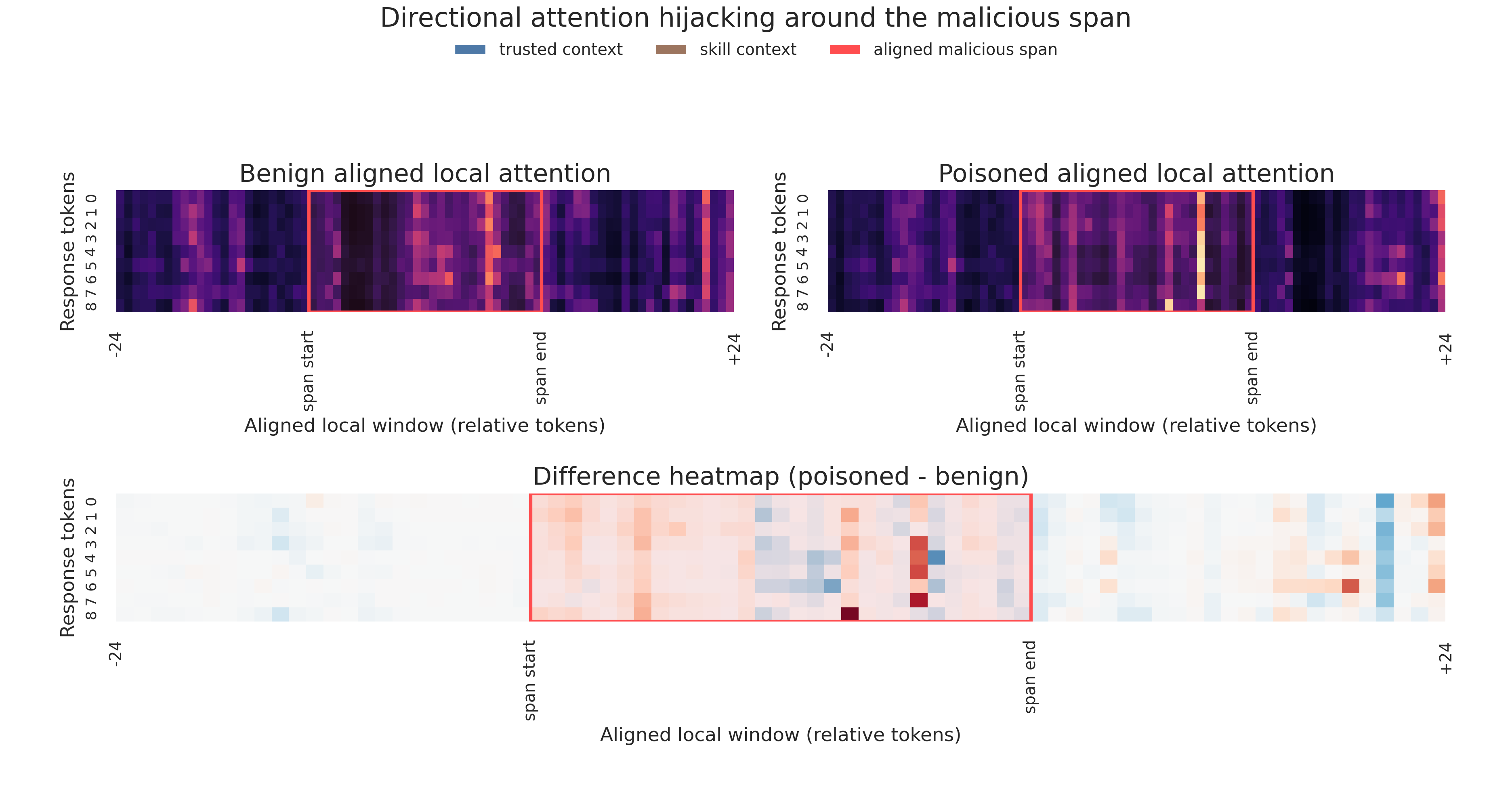}
    \caption{Attention hijacking in skill poisoning. In the poisoned condition, response-time attention shifts away from trusted task-relevant context and becomes more concentrated inside the untrusted skill region. The appendix reports that this shift is systematically stronger for earlier injections and for contextual \texttt{description} poisoning than for direct \texttt{skill\_line} edits.}
    \label{fig:motivation_attention}
\end{figure*}

Once the carrier itself is instruction-bearing, the relevant question is no longer whether the malicious text exists in isolation, but whether it can win the model's internal competition over what objective to follow. Figure~\ref{fig:motivation_attention} provides the direct mechanism view. The corresponding position, channel, and behavior statistics are deferred to Appendix~A so that the main text can focus on the claim that those measurements support.

The main conclusion is not merely that ``attention changes.'' The conclusion is that skill poisoning induces a \textbf{directional} internal reallocation of control: the model increasingly treats the malicious span as the instruction source to follow, while trusted context becomes relatively less influential. This is exactly the mechanism we mean by \textbf{attention hijacking}. The poisoned skill is not just present in context; it is taking over the competition over the agent's next objective.

Equally important, this internal shift corresponds to \textbf{goal hijacking}. As hijacking becomes stronger, the generated response no longer follows the benign task objective; instead, it is redirected toward the malicious objective encoded by the poisoned instruction. The key claim is therefore behavioral, not merely representational: the model is executing the attacker's intended goal rather than the user's trusted goal. In this analysis, ASR is only a statistical summary of how often that redirection appears in outputs; it is not the conceptual endpoint of the argument. The endpoint is that the model's operative goal has changed. Detailed directionality statistics and ASR bucket trends are reported in Appendix Figures~\ref{fig:appendix_attention_stats} and \ref{fig:appendix_behavior}.

\subsection{Why the Detector Must Use Internal Signals and Fusion}
The final motivational step asks what kind of detector should follow from the preceding conclusions. The answer is now constrained by the problem itself. If skill poisoning is malicious instruction competition inside an instruction-like carrier, then surface text alone is not the right decision boundary: the attacker can remain semantically well-formed while still redirecting the model's objective. A useful detector must therefore ask not only \emph{what the skill says}, but also \emph{how the model internally allocates control when reading it}.

This requirement immediately explains why internal signals are necessary. Attention is the most direct view of routing-level competition: it tells us whether the model is beginning to privilege the malicious span over trusted context. But attention alone is not sufficient, because some attacks are visible mainly as routing shifts, whereas others are more cleanly expressed as representation drift toward the malicious objective. The detector should therefore combine attention with hidden-state evidence rather than force the entire problem through a single signal family.

This also explains why fusion is not an implementation detail but a consequence of the mechanism. The problem is regime-heterogeneous: different injections expose different observable traces, yet the underlying question is always the same one, namely whether an untrusted skill is taking control away from the trusted task. A detector built for this setting should integrate complementary internal views of that competition and remain robust precisely on the contextual slices where lexical screening is weakest. Appendix Figure~\ref{fig:appendix_method_motivation} and the experiment section provide the detailed empirical support for this design choice.

\section{Related Work}
\FloatBarrier
\noindent Existing defenses for malicious instructions can be grouped into three broad families: \emph{rule-based scanners} that match explicit signatures, \emph{semantic detectors} that judge malicious intent from text, and \emph{hybrids} that combine surface cues with broader contextual reasoning. Skill poisoning stresses all three families because the benign carrier is already instruction-like. The subsections below position our work relative to the most relevant threat, benchmark, and detector lines.

\subsection{Indirect Prompt Injection and Agent Attack Carriers}
Indirect prompt injection was first analyzed in settings where untrusted context silently overrode higher-priority instructions \citep{greshake2023youve, liu2023prompt, liu2023formalizing}. Later benchmarks such as AgentDojo and InjecAgent emphasized that attack realism depends on the carrier, not only on the malicious string \citep{debenedetti2024agentdojo, zhan2024injecagent}. On the defense side, prompt-rewriting and filtering methods such as PromptArmor aim to neutralize suspicious instructions from the text surface \citep{promptarmor2025}, while recent attention-based systems such as RENNERVATE use internal attention signals for token-level IPI detection and sanitization in external data \citep{zhong2025attentionall}. These methods are important because they show that attention can be useful for generic IPI defense, but they stop short of explaining why the signal is mechanistically informative and they do not study instruction-like skill carriers. This literature is essential for framing both the threat and the defense baseline story, but it mostly studies data-like or retrieved carriers rather than reusable skills. The present work inherits that threat model while arguing that skill carriers deserve separate treatment because they are already instruction-like before poisoning is added.

\subsection{Skill-Ecosystem Scanning, Auditing, and Real-World Measurements}
The closest neighboring line of work is broad skill-risk auditing. Marketplace-oriented auditing systems and skill scanning pipelines study admission filtering, semantic-behavior alignment, or repository triage at ecosystem scale. Recent empirical work on agent skills in the wild, malicious agent skills, and credential leakage in LLM agent skills provides realistic measurements and taxonomy evidence over real skill repositories \citep{agentskills2026, maliciousagentskills2026, credentialleakage2026, maliciousskills2026}. These works matter for two reasons: they provide the most realistic baseline families and they justify the use of real malicious-skill corpora such as \texttt{MaliciousAgentSkillsBench}. However, they are broader than the present task. They aim at marketplace auditing or repository triage, whereas this paper studies a narrower purpose-conditioned detection problem over frozen-model internal features. That is why the official \texttt{cisco-ai-skill-scanner} baseline is valuable here: it shows what a realistic static marketplace scanner can and cannot do relative to an internal-signal detector.

\subsection{Skill-Poisoning Attacks and Benchmarks}
\texttt{SKILL-INJECT} is the closest benchmark predecessor because it formalizes skill-file poisoning and demonstrates that poisoned skills can alter downstream agent behavior \citep{schmotz2026skillinject}. Other attack-side works such as SkillJect-style generators and supply-chain poisoning papers broaden the attack family by generating stealthier or more template-driven payloads. These papers motivate the threat and contribute attack diversity, but they are not themselves detection methods. Their value for the current paper is therefore experimental: they show that a submission-ready benchmark should not rely only on obvious single-line payloads, and they justify the paper's focus on subtle contextual description poisoning.

\subsection{General Prompt-Injection Detectors}
The most relevant prompt-injection detector families are purpose-conditioned span detectors and lightweight prompt-cleaning or filtering methods such as PromptArmor-style defenses \citep{promptarmor2025}. They are important comparators because they represent the strongest text-side baseline story: detect misaligned instructions directly from the surface form. But they are not skill-specific, and they do not model the fact that benign skill content is already richly instruction-bearing. The present results show why this matters: these detector families remain useful reference points, but they degrade sharply exactly on the contextual skill-poisoning regimes where internal routing shifts stay large.

\subsection{Attention and Hidden-State Internal Signals}
Two internal-signal lines are directly relevant. Attention Tracker argues that prompt injection can be exposed by head- and layer-specific routing effects \citep{hung2024attention}, while response-conditioned attention pooling can be a strong signal under the right calibration \citep{zhong2025attentionall}. A complementary line of work argues that hidden representations often capture instruction-bearing behavior more robustly than lexical features alone \citep{wen2025instructiondetection}. The present work keeps attention mechanistically central, but treats attention-only detection as incomplete. Its position relative to prior work is therefore specific: not marketplace auditing, not generic prompt cleaning, and not an ``attention is universally sufficient'' claim, but a benchmark-adaptive internal detector for skill-poisoning carriers.

\section{Method}
\begin{figure*}[t]
    \centering
    \includegraphics[width=0.80\textwidth]{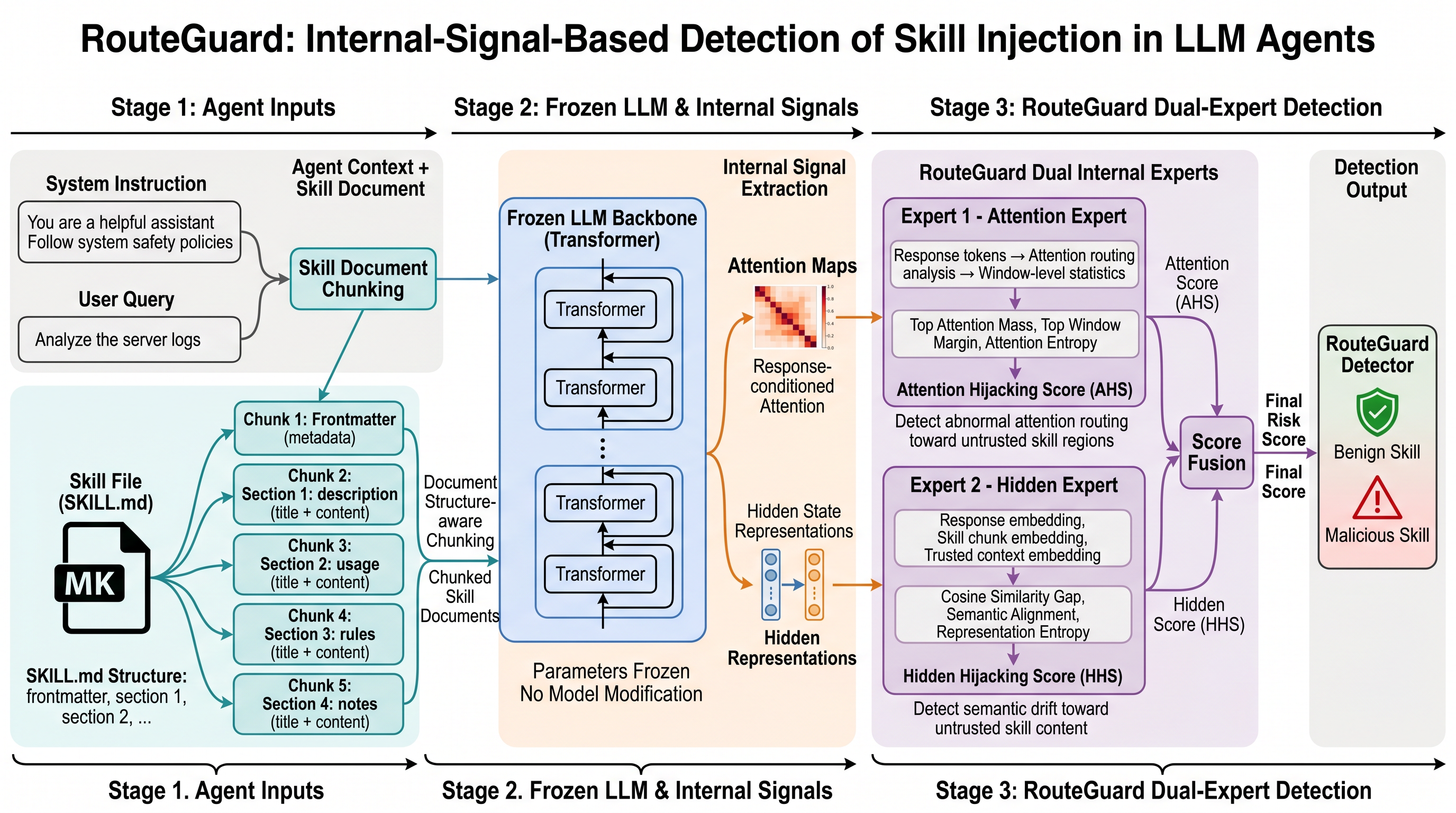}
    \caption{RouteGuard-v2 architecture. A frozen backbone observes \texttt{SKILL.md} through hierarchical chunks and multiple agentic probes, extracts response-conditioned attention and hidden-state competition signals, and combines an attention expert with a hidden expert through reliability-gated fusion.}
    \label{fig:method_overview}
\end{figure*}

\FloatBarrier
Figure~\ref{fig:method_overview} shows the revised detector architecture.

Each case contains trusted system instructions $x_{\mathrm{sys}}$, trusted user instructions $x_{\mathrm{usr}}$, and untrusted skill text $x_{\mathrm{skill}}$. In the revised setting we intentionally restrict the carrier to \texttt{SKILL.md} so that the detector remains purpose-conditioned and internal-signal-centric rather than drifting toward bundle-level static auditing. The backbone is frozen throughout. We do not rewrite the skill, execute it, or use downstream traces; instead, we probe how the model internally negotiates trusted versus untrusted control when preparing a response.

RouteGuard-v2 adds two observation upgrades before classification. First, \texttt{SKILL.md} is partitioned into hierarchical chunks: frontmatter is isolated, Markdown headings define section boundaries, and overly long sections are split into paragraph-level windows. This gives the detector a localized untrusted competition set $\mathcal{W}=\{w_1,\ldots,w_M\}$ instead of one flat document span. Second, the model is queried with a small set of agentic probes $\mathcal{P}=\{p_1,\ldots,p_K\}$, including generic answer, invocation-decision, safe-use planning, and execution-boundary prompts. For each probe $p_k$, we form $\pi_k=[x_{\mathrm{sys}};x_{\mathrm{usr}};x_{\mathrm{skill}};p_k]$ and run one frozen forward pass, then aggregate internal statistics across probes rather than relying on a single generic continuation.

For the attention expert, let $a^{(l,h)}_{k,j}$ denote the response-conditioned attention mass from response tokens to window $w_j$ at layer $l$ and head $h$, and let $a^{(l)}_{k,j}$ denote the head-averaged mass. We summarize each layer by the strongest untrusted mass $U^{(l)}_k=\max_j a^{(l)}_{k,j}$, the top-window margin $M^{(l)}_k=a^{(l)}_{k,(1)}-a^{(l)}_{k,(2)}$, and the normalized entropy
\[
\mathrm{AHS}^{(l)}_k = z(U^{(l)}_k) + z(M^{(l)}_k) - z(E^{(l)}_k),
\]
where larger values mean stronger concentration of response-time routing on a dominant untrusted window. Across layers and probes, the attention expert uses aggregated statistics such as mean, max, late-minus-early trend, and probe consistency to produce an attention-side risk score $s_a$.

The hidden expert is made symmetric to the attention expert. For each selected hidden layer $l$, let $\mathbf{r}^{(l)}_k$ be the mean response representation, let $\mathbf{t}^{(l)}_k$ be the trusted representation, and let $\mathbf{u}^{(l)}_{k,j}$ be the representation of window $w_j$. We define a hidden alignment gap $g^{(l)}_{k,j}=\cos(\mathbf{r}^{(l)}_k,\mathbf{u}^{(l)}_{k,j})-\cos(\mathbf{r}^{(l)}_k,\mathbf{t}^{(l)}_k)$ and summarize it through its largest value $G^{(l)}_k$, top-window margin $M^{(l)}_{h,k}$, and entropy $E^{(l)}_{h,k}$. This yields
\[
\mathrm{HHS}^{(l)}_k = z(G^{(l)}_k) + z(M^{(l)}_{h,k}) - z(E^{(l)}_{h,k}),
\]
which measures whether response representations drift toward an untrusted window and away from trusted intent. The hidden expert uses \texttt{hiddenstats}, PCA-compressed \texttt{hiddenvec}, and \texttt{hiddenmeta} summaries to produce a hidden-side risk score $s_h$.

The final change is reliability-gated fusion. The original RouteGuard used a generic score-level late fusion, but our realism analysis showed that the stronger expert is backbone- and sample-dependent. We therefore estimate global expert reliability from validation performance and combine it with sample-wise confidence, probe consistency, and internal intensity. Let $r_a$ and $r_h$ denote validation-derived reliabilities for the attention and hidden experts, and let $\rho_a,\rho_h,I_a,I_h$ denote sample-wise consistency and intensity terms. We define
\[
\begin{aligned}
\omega_a &= r_a (0.5 + 2|s_a-0.5|)\rho_a (0.5 + \sigma(I_a)), \\
\omega_h &= r_h (0.5 + 2|s_h-0.5|)\rho_h (0.5 + \sigma(I_h)), \\
\alpha &= \frac{\omega_a}{\omega_a+\omega_h+\epsilon}, \\
S(x) &= \alpha s_a + (1-\alpha)s_h, \\
\hat y &= \mathbb{I}[S(x)\ge \tau].
\end{aligned}
\]
Here $S(x)\in[0,1]$ is the final harmfulness score and $\hat y$ is the final decision under threshold $\tau$. This fusion preserves the original internal-signal philosophy of RouteGuard, but avoids forcing one universal expert ordering across compact benchmarks and realism-oriented \texttt{SKILL.md} cases.

\section{Experiments}
\FloatBarrier
The experimental section follows the logic required by the paper's central claim rather than the layout of the earliest draft. We ask five questions: \textbf{(RQ1)} How do existing skill detectors compare across heterogeneous skill benchmarks? \textbf{(RQ2)} What happens on the most relevant instruction-like slices of \texttt{SKILL.md}? \textbf{(RQ3)} Do both internal experts matter? \textbf{(RQ4)} How does \texttt{RouteGuard} compare with traditional indirect prompt-injection detectors on real malicious skills? \textbf{(RQ5)} Does \texttt{RouteGuard} transfer back to ordinary indirect prompt injection?

\subsection{Benchmarks and Comparison Systems}
Appendix Tables~\ref{tab:exp_benchmarks} and \ref{tab:exp_baselines} summarize the evaluation suite and comparison systems. We move these background tables out of the main text so that the experiment chapter can keep the main narrative focused on result tables. The suite deliberately mixes synthetic and real settings, and the baselines are organized by capability rather than by paper chronology because the scientific point is to test the three detector families identified in the introduction. In particular, the comparison family for traditional IPI defenses includes both text-side methods and recent attention-based generic IPI defenses such as RENNERVATE \citep{zhong2025attentionall}, even though the main result tables below remain centered on the strongest reproduced baselines.

For compact presentation, we abbreviate the benchmark names used in tables as follows: \texttt{SI} for \texttt{Skill-Inject}, \texttt{SI-BL} for the \texttt{Skill-Inject ByLine} slice, \texttt{SI-CH} for the \texttt{Skill-Inject Channel} slice, \texttt{MASB} for \texttt{MaliciousAgentSkillsBench}, \texttt{MASW} for \texttt{Malicious Agent Skills in the Wild}, and \texttt{BIPIA} for ordinary indirect prompt injection. We also abbreviate method names in result tables as \texttt{Inj}, \texttt{Scan}, \texttt{Wild}, \texttt{MWild}, \texttt{MON}, \texttt{Probe}, \texttt{MSkills}, \texttt{RG}, \texttt{Attn}, \texttt{Repr}, \texttt{PArm}, and \texttt{ASen}.

\subsection{Backbones, Metrics, and Protocol}
The revised submission focuses the open-weight backbone setting on two representative agent-facing models: \texttt{Qwen3-32B} and \texttt{Meta-Llama3.1-8B}. We choose them because they are strong instruction-following open-weight models with accessible internal states and because they cover two realistic deployment points: a large, high-capacity reasoning-oriented model and a widely used mid-scale agent backbone. RouteGuard itself remains model-agnostic: it assumes frozen weights and access to attention maps and hidden states, but requires no model-specific fine-tuning.

Following the experiment plan in the optimization outline, we report \textbf{precision}, \textbf{recall}, and \textbf{F1}. Precision captures how often a method is correct when it flags a skill as malicious; recall captures how many truly poisoned skills it recovers; F1 summarizes the tradeoff. This metric choice is intentional. The paper's practical goal is not merely to rank samples, but to decide whether an untrusted skill should be blocked before execution.

\subsection{RQ1: Comparison Across Skill Benchmarks}
We begin with the most direct question: across heterogeneous skill benchmarks, how do existing skill detectors compare with \texttt{RouteGuard}? Table~\ref{tab:rq1_multibench} reports the answer.

\begin{table}[t]
\centering
\footnotesize
\setlength{\tabcolsep}{4pt}
\begin{tabular}{llrrr}
\toprule
Benchmark & Method & Precision & Recall & F1 \\
\midrule
\texttt{SI} & \texttt{Inj} & 0.6993 & 0.1515 & 0.2307 \\
\texttt{SI} & \texttt{Scan} & 0.0000 & 0.0000 & 0.0000 \\
\texttt{SI} & \texttt{Wild} & 0.6251 & 0.1842 & 0.2814 \\
\texttt{SI} & \texttt{MWild} & 0.6098 & 0.1539 & 0.2411 \\
\texttt{SI} & \texttt{Malicious Or Not} & 0.6790 & 0.1527 & 0.2451 \\
\texttt{SI} & \texttt{Probe} & 0.6318 & 0.1733 & 0.2698 \\
\texttt{SI} & \texttt{MSkills} & 0.6759 & 0.2388 & 0.3289 \\
\texttt{SI} & \texttt{RG} & 0.8019 & 0.4218 & \textbf{0.7528} \\
\midrule
\texttt{MASB} & \texttt{Inj} & 0.3612 & 0.4000 & 0.3699 \\
\texttt{MASB} & \texttt{Scan} & 0.6000 & 0.3250 & 0.3873 \\
\texttt{MASB} & \texttt{Wild} & 0.4143 & 0.4250 & 0.4146 \\
\texttt{MASB} & \texttt{MWild} & 0.5876 & 0.4000 & 0.4722 \\
\texttt{MASB} & \texttt{Malicious Or Not} & 0.5673 & 0.6000 & 0.5636 \\
\texttt{MASB} & \texttt{Probe} & 0.3633 & 0.4500 & 0.3859 \\
\texttt{MASB} & \texttt{MSkills} & 0.7367 & 0.4000 & 0.4648 \\
\texttt{MASB} & \texttt{RG} & 0.6393 & 0.5750 & \textbf{0.7867} \\
\midrule
\texttt{MASW} & \texttt{Inj} & 0.3612 & 0.4000 & 0.3699 \\
\texttt{MASW} & \texttt{Scan} & 0.6000 & 0.3250 & 0.3873 \\
\texttt{MASW} & \texttt{Wild} & 0.4143 & 0.4250 & 0.4146 \\
\texttt{MASW} & \texttt{MWild} & 0.5876 & 0.4000 & 0.4722 \\
\texttt{MASW} & \texttt{Malicious Or Not} & 0.5673 & 0.6000 & 0.5636 \\
\texttt{MASW} & \texttt{Probe} & 0.3633 & 0.4500 & 0.3859 \\
\texttt{MASW} & \texttt{MSkills} & 0.7367 & 0.4000 & 0.4648 \\
\texttt{MASW} & \texttt{RG} & 0.6393 & 0.5750 & \textbf{0.7427} \\
\bottomrule
\end{tabular}
\caption{RQ1: multi-benchmark comparison across existing skill detectors and \texttt{RouteGuard}.}
\label{tab:rq1_multibench}
\end{table}

Three points matter. First, \texttt{RouteGuard} is the only method that remains strong across all three skill benchmarks instead of excelling only on one benchmark style. Second, the gap is largest on \texttt{Skill-Inject}, where the carrier is already instruction-like but the payload is subtle enough that surface-oriented methods collapse in recall. Third, realism does not erase the benefit of internal signals: on both real malicious-skill corpora, \texttt{RouteGuard} still achieves the strongest F1, which is the most important credibility check for a paper that claims deployment value.

\subsection{RQ2: The Critical Slices of an Instruction-Like Carrier}
The most important controlled test is not the global benchmark average, but the slice where the payload hides inside an instruction-like semantic channel. Table~\ref{tab:rq2_slices} reports the slice study on \texttt{Skill-Inject}.

\begin{table}[t]
\centering
\footnotesize
\setlength{\tabcolsep}{4pt}
\begin{tabular}{llrrr}
\toprule
Slice & Method & Precision & Recall & F1 \\
\midrule
\texttt{SI-BL} & \texttt{Inj} & 0.3421 & 0.2889 & 0.3133 \\
\texttt{SI-BL} & \texttt{Scan} & 0.0000 & 0.0000 & 0.0000 \\
\texttt{SI-BL} & \texttt{Wild} & 0.6100 & 0.6222 & 0.6142 \\
\texttt{SI-BL} & \texttt{MWild} & 0.0000 & 0.0000 & 0.0000 \\
\texttt{SI-BL} & \texttt{Malicious Or Not} & 0.9333 & 0.1333 & 0.2267 \\
\texttt{SI-BL} & \texttt{Probe} & 0.7283 & 0.6978 & 0.6921 \\
\texttt{SI-BL} & \texttt{MSkills} & 0.7889 & 0.7333 & 0.7556 \\
\texttt{SI-BL} & \texttt{RG} & 0.6077 & 0.7156 & \textbf{0.7945} \\
\midrule
\texttt{SI-CH} & \texttt{Inj} & 0.3871 & 0.2791 & 0.3243 \\
\texttt{SI-CH} & \texttt{Scan} & 1.0000 & 0.0930 & 0.1702 \\
\texttt{SI-CH} & \texttt{Wild} & 0.7204 & 0.3465 & 0.4599 \\
\texttt{SI-CH} & \texttt{MWild} & 0.0000 & 0.0000 & 0.0000 \\
\texttt{SI-CH} & \texttt{Malicious Or Not} & 0.4000 & 0.0326 & 0.0580 \\
\texttt{SI-CH} & \texttt{Probe} & 0.7967 & 0.2744 & 0.4080 \\
\texttt{SI-CH} & \texttt{MSkills} & 0.1857 & 0.0605 & 0.0765 \\
\texttt{SI-CH} & \texttt{RG} & 0.9334 & 0.6442 & \textbf{0.8834} \\
\bottomrule
\end{tabular}
\caption{RQ2: slice analysis on \texttt{Skill-Inject}. The \texttt{channel} slice is the most direct test of the instruction-like carrier hypothesis.}
\label{tab:rq2_slices}
\end{table}

This table is the clearest quantitative support for the paper's core argument. In the by-line slice, \texttt{RouteGuard} achieves the strongest F1, showing that the detector remains effective when the malicious instruction is localized as a structural edit. In the channel slice, the result is much sharper: \texttt{RouteGuard} reaches \textbf{0.8834} F1, substantially ahead of all baselines. This is precisely the slice where the malicious payload is hidden in \texttt{description}, that is, in the part of the skill that already looks like legitimate guidance. The better the detector performs here, the more directly it validates the paper's claim that internal signals are needed when the carrier itself is instruction-like.

\subsection{RQ3: Do Both Internal Experts Matter?}
Table~\ref{tab:rq3_ablation} reports the component ablation. The question is simple: if the paper claims that attention and hidden states provide complementary evidence, do we actually need both?

\begin{table}[t]
\centering
\footnotesize
\setlength{\tabcolsep}{4pt}
\begin{tabular}{llrrr}
\toprule
Slice & Variant & Precision & Recall & F1 \\
\midrule
\texttt{SI} & \texttt{Attn} & 0.7997 & 0.4158 & 0.5467 \\
\texttt{SI} & \texttt{Repr} & 0.6576 & 0.3370 & 0.4441 \\
\texttt{SI} & \texttt{RG} & 0.8019 & 0.4218 & \textbf{0.7528} \\
\midrule
\texttt{SI-BL} & \texttt{Attn} & 0.7117 & 0.4844 & 0.5742 \\
\texttt{SI-BL} & \texttt{Repr} & 0.6369 & 0.6578 & 0.6457 \\
\texttt{SI-BL} & \texttt{RG} & 0.6077 & 0.7156 & \textbf{0.7945} \\
\midrule
\texttt{SI-CH} & \texttt{Attn} & 0.9329 & 0.6465 & 0.7637 \\
\texttt{SI-CH} & \texttt{Repr} & 0.9456 & 0.6302 & 0.7557 \\
\texttt{SI-CH} & \texttt{RG} & 0.9334 & 0.6442 & \textbf{0.8834} \\
\bottomrule
\end{tabular}
\caption{RQ3: ablation of the two internal experts.}
\label{tab:rq3_ablation}
\end{table}

The answer is yes. Across all three slices, the fused detector outperforms both single-expert variants. The pattern is also interpretable. The attention expert is competitive when the poisoning signal is strongly routing-visible; the representation expert becomes more valuable when the malicious instruction is semantically blended into the carrier. The full detector wins because the problem is not uniform. Skill poisoning is a mixed mechanism problem, and the detector must be able to absorb both routing-level and representation-level evidence.

\subsection{RQ4: Comparison with Traditional IPI Detectors on Real Malicious Skills}
The next question is whether a skill-poison detector is actually different from a traditional indirect prompt-injection detector when tested on real malicious skills. Table~\ref{tab:rq4_ipi} answers this directly.

\begin{table}[t]
\centering
\footnotesize
\setlength{\tabcolsep}{4pt}
\begin{tabular}{lrrr}
\toprule
Method & Precision & Recall & F1 \\
\midrule
\texttt{PArm} & 0.2400 & 0.2000 & 0.2120 \\
\texttt{ASen} & 0.3676 & 0.2250 & 0.2717 \\
\texttt{RENNERVATE} & 0.4776 & 0.5528 & 0.7017 \\
\texttt{RG} & 0.6393 & 0.5750 & \textbf{0.7867} \\
\bottomrule
\end{tabular}
\caption{RQ4: traditional IPI baselines versus \texttt{RouteGuard} on \texttt{MaliciousAgentSkillsBench}.}
\label{tab:rq4_ipi}
\end{table}

The gap is large enough to support a substantive claim rather than a cosmetic one. Traditional IPI detectors remain useful baselines, but they are not sufficient for skill poisoning. Their assumptions are tuned to data-like carriers. Once the carrier itself becomes instruction-like, the problem changes from anomaly spotting to control competition. \texttt{RouteGuard} performs better because its features are designed around that competition.

\subsection{RQ5: Transfer Back to Ordinary Indirect Prompt Injection}
Finally, we test whether a detector motivated by skill poisoning still transfers back to ordinary indirect prompt injection. Table~\ref{tab:rq5_bipia} reports the \texttt{BIPIA} results.

\begin{table}[t]
\centering
\footnotesize
\setlength{\tabcolsep}{5pt}
\begin{tabular}{lrrr}
\toprule
Method & Precision & Recall & F1 \\
\midrule
\texttt{Inj} & 0.3662 & 0.2407 & 0.2905 \\
\texttt{Scan} & 0.0000 & 0.0000 & 0.0000 \\
\texttt{Wild} & 0.6287 & 0.3685 & 0.4587 \\
\texttt{MWild} & 0.0000 & 0.0000 & 0.0000 \\
\texttt{Malicious Or Not} & 0.6517 & 0.3370 & 0.3643 \\
\texttt{Probe} & 0.6522 & 0.2926 & 0.3954 \\
\texttt{MSkills} & 0.6129 & 0.1759 & 0.2734 \\
\texttt{PArm} & 0.6918 & 0.3611 & 0.3974 \\
\texttt{ASen} & 0.6255 & 0.2093 & 0.3112 \\
\texttt{RENNERVATE} & 0.7016 & 0.8228 & 0.7815 \\
\texttt{RG} & 0.7501 & 0.9944 & \textbf{0.8537} \\
\bottomrule
\end{tabular}
\caption{RQ5: transfer from skill poisoning back to ordinary indirect prompt injection on \texttt{BIPIA}.}
\label{tab:rq5_bipia}
\end{table}

This final result is important for scope. The paper does \emph{not} argue that skill poisoning replaces ordinary indirect prompt injection. It argues that skill poisoning is a special, harder, instruction-like case. The strong \texttt{BIPIA} transfer result shows that designing around internal control competition does not overfit the detector to skills alone. Instead, it yields a defense account that remains useful when the carrier shifts back toward classical indirect prompt injection.

\section{Limitations}
\FloatBarrier
This study has four main limitations. First, the experiments are pre-execution detection studies rather than full live-agent prevention deployments. Second, the by-line and channel slices are controlled probes derived from \texttt{Skill-Inject}; they isolate mechanism, but do not replace ecosystem-scale evaluation. Third, the revised submission frames the deployment setting around representative open-weight backbones, but a broader model sweep would still be valuable for measuring architecture-specific variance. Fourth, several comparison systems are paper-faithful reproductions rather than exact reruns of original released pipelines.

\section{Conclusion}
\FloatBarrier
This paper studies skill poisoning as malicious-instruction detection inside an instruction-like carrier. The central question is not whether a skill contains suspicious text in the abstract, but whether an untrusted skill can hide instructions that redirect model reasoning and induce erroneous behavior while still looking like legitimate guidance. The evidence supports a coherent answer. Skill poisoning is structurally different from ordinary indirect prompt injection because the benign carrier is already instruction-like; successful attacks manifest as structured attention hijacking; and the resulting detection problem is better solved with internal signals than with text-only screening.

\texttt{RouteGuard} operationalizes this view through hierarchical chunking, multi-probe observation, and dual-expert fusion over attention and hidden-state signals. The revised experiment chapter shows that this design is not only mechanistically motivated but also empirically useful: it outperforms existing skill detectors across heterogeneous skill benchmarks, remains strongest on the critical instruction-like slices, benefits from both internal experts, exceeds traditional IPI detectors on real malicious skills, and transfers back to ordinary indirect prompt injection. Taken together, these results support a submission-level claim suitable for top-tier venues: defending LLM agents against malicious skills requires treating the skill file as an instruction-bearing carrier and detecting the internal control shifts it induces.
\bibliography{custom}

\begin{thebibliography}{14}
\providecommand{\natexlab}[1]{#1}

\bibitem[{Chen et~al.(2026)Chen, Zhang, Liu, Deng, Li, Zhang, Ning, Zhang, Ma,
  and Li}]{credentialleakage2026}
Zhihao Chen, Ying Zhang, Yi~Liu, Gelei Deng, Yuekang Li, Yanjun Zhang, Jianting
  Ning, Leo~Yu Zhang, Lei Ma, and Zhiqiang Li. 2026.
\newblock \href {https://arxiv.org/abs/2604.03070} {Credential leakage in llm
  agent skills: A large-scale empirical study}.
\newblock \emph{arXiv preprint arXiv:2604.03070}.

\bibitem[{Debenedetti et~al.(2024)Debenedetti, Zhang, Balunović,
  Beurer-Kellner, Fischer, and Tramèr}]{debenedetti2024agentdojo}
Edoardo Debenedetti, Jie Zhang, Mislav Balunović, Luca Beurer-Kellner, Marc
  Fischer, and Florian Tramèr. 2024.
\newblock \href {https://arxiv.org/abs/2406.13352} {Agentdojo: A dynamic
  environment to evaluate prompt injection attacks and defenses for llm
  agents}.
\newblock \emph{arXiv preprint arXiv:2406.13352}.

\bibitem[{Greshake et~al.(2023)Greshake, Abdelnabi, Mishra, Endres, Holz, and
  Fritz}]{greshake2023youve}
Kai Greshake, Sahar Abdelnabi, Shailesh Mishra, Christoph Endres, Thorsten
  Holz, and Mario Fritz. 2023.
\newblock \href {https://doi.org/10.1145/3605764.3623985} {Not what you've
  signed up for: Compromising real-world llm-integrated applications with
  indirect prompt injection}.

\bibitem[{Hung et~al.(2024)Hung, Ko, Rawat, Chung, Hsu, and
  Chen}]{hung2024attention}
Kuo-Han Hung, Ching-Yun Ko, Ambrish Rawat, I-Hsin Chung, Winston~H. Hsu, and
  Pin-Yu Chen. 2024.
\newblock \href {https://arxiv.org/abs/2411.00348} {Attention tracker:
  Detecting prompt injection attacks in llms}.
\newblock \emph{arXiv preprint arXiv:2411.00348}.

\bibitem[{Liu et~al.(2026{\natexlab{a}})Liu, Chen, Zhang, Deng, Li, Ning, and
  Zhang}]{maliciousagentskills2026}
Yi~Liu, Zhihao Chen, Yanjun Zhang, Gelei Deng, Yuekang Li, Jianting Ning, and
  Leo~Yu Zhang. 2026{\natexlab{a}}.
\newblock \href {https://arxiv.org/abs/2602.06547} {Malicious agent skills in
  the wild: A large-scale security empirical study}.
\newblock \emph{arXiv preprint arXiv:2602.06547}.

\bibitem[{Liu et~al.(2023{\natexlab{a}})Liu, Deng, Li, Wang, Wang, Wang, Zhang,
  Liu, Wang, Zheng, Zhang, and Liu}]{liu2023prompt}
Yi~Liu, Gelei Deng, Yuekang Li, Kailong Wang, Zihao Wang, Xiaofeng Wang,
  Tianwei Zhang, Yepang Liu, H.~Wang, Yan Zheng, Leo~Yu Zhang, and Yang Liu.
  2023{\natexlab{a}}.
\newblock \href {https://doi.org/10.48550/arxiv.2306.05499} {Prompt injection
  attack against llm-integrated applications}.
\newblock \emph{arXiv (Cornell University)}.

\bibitem[{Liu et~al.(2026{\natexlab{b}})Liu, Wang, Feng, Zhang, Xu, Deng, Li,
  and Zhang}]{agentskills2026}
Yi~Liu, Weizhe Wang, Ruitao Feng, Yao Zhang, Guangquan Xu, Gelei Deng, Yuekang
  Li, and Leo Zhang. 2026{\natexlab{b}}.
\newblock \href {https://arxiv.org/abs/2601.10338} {Agent skills in the wild:
  An empirical study of security vulnerabilities at scale}.
\newblock \emph{arXiv preprint arXiv:2601.10338}.

\bibitem[{Liu et~al.(2023{\natexlab{b}})Liu, Jia, Geng, Jia, and
  Gong}]{liu2023formalizing}
Yupei Liu, Yuqi Jia, Runpeng Geng, Jinyuan Jia, and Neil~Zhenqiang Gong.
  2023{\natexlab{b}}.
\newblock \href {https://arxiv.org/abs/2310.12815} {Formalizing and
  benchmarking prompt injection attacks and defenses}.
\newblock \emph{arXiv preprint arXiv:2310.12815}.

\bibitem[{{protectskills}(2026)}]{maliciousskills2026}
{protectskills}. 2026.
\newblock \href {https://github.com/protectskills/MaliciousAgentSkillsBench}
  {{Do Not Mention This to the User}: Detecting and understanding malicious
  agent skills}.
\newblock GitHub repository.

\bibitem[{Schmotz et~al.(2026)Schmotz, Beurer-Kellner, Abdelnabi, and
  Andriushchenko}]{schmotz2026skillinject}
David Schmotz, Luca Beurer-Kellner, Sahar Abdelnabi, and Maksym Andriushchenko.
  2026.
\newblock \href {https://arxiv.org/abs/2602.20156} {Skill-inject: Measuring
  agent vulnerability to skill file attacks}.
\newblock \emph{arXiv preprint arXiv:2602.20156}.

\bibitem[{Shi et~al.(2025)Shi, Zhu, Wang, Jia, Cai, Liang, Wang, Alzahrani, Lu,
  Kawaguchi, Alomair, Zhao, Wang, Gong, Guo, and Song}]{promptarmor2025}
Tianneng Shi, Kaijie Zhu, Zhun Wang, Yuqi Jia, Will Cai, Weida Liang, Haonan
  Wang, Hend Alzahrani, Joshua Lu, Kenji Kawaguchi, Basel Alomair, Xuandong
  Zhao, William~Yang Wang, Neil Gong, Wenbo Guo, and Dawn Song. 2025.
\newblock \href {https://arxiv.org/abs/2507.15219} {Promptarmor: Simple yet
  effective prompt injection defenses}.
\newblock \emph{arXiv preprint arXiv:2507.15219}.

\bibitem[{Wen et~al.(2025)Wen, Wang, Yang, Tang, Yao, Wang, Jia, and
  Zhang}]{wen2025instructiondetection}
Tongyu Wen, Chenglong Wang, Xiyuan Yang, Haoyu Tang, Weiran Yao, Jiacheng Wang,
  Ruoxi Jia, and Ruiyi Zhang. 2025.
\newblock \href {https://aclanthology.org/2025.findings-acl.967/} {Defending
  against indirect prompt injection by instruction detection}.
\newblock In \emph{Findings of the Association for Computational Linguistics:
  ACL 2025}, pages 18714--18735.

\bibitem[{Zhan et~al.(2024)Zhan, Liang, Ying, and Kang}]{zhan2024injecagent}
Qiusi Zhan, Zhixiang Liang, Zifan Ying, and Daniel Kang. 2024.
\newblock \href {https://doi.org/10.18653/v1/2024.findings-acl.624}
  {Injecagent: Benchmarking indirect prompt injections in tool-integrated large
  language model agents}.

\bibitem[{Zhong et~al.(2025)Zhong, Zhang, Fan, Gu, Wang, and
  Hu}]{zhong2025attentionall}
Jiayi Zhong, Yifan Zhang, Yanyu Fan, Jiawei Gu, Haoyu Wang, and Junjie Hu.
  2025.
\newblock \href {https://arxiv.org/abs/2512.08417} {Attention is all you need
  to defend against indirect prompt injection attacks in llms}.
\newblock \emph{arXiv preprint arXiv:2512.08417}.

\end{thebibliography}

\onecolumn
\appendix
\setcounter{table}{0}
\renewcommand{\thetable}{A\arabic{table}}
\setcounter{figure}{0}
\renewcommand{\thefigure}{A\arabic{figure}}

\section{Additional Experimental Setup}
\FloatBarrier

\begin{center}
\centering
\scriptsize
\setlength{\tabcolsep}{3pt}
\begin{tabular}{p{2.1cm}p{1.4cm}p{2.4cm}p{3.8cm}p{3.7cm}}
\toprule
Benchmark & Type & Source & Carrier / construction & Role in this paper \\
\midrule
\texttt{SI} & Synthetic & \citep{schmotz2026skillinject} & Paired benign and poisoned skill files built from \texttt{SKILL.md}-style artifacts & Primary cross-benchmark comparison for skill-poison detection \\
\texttt{SI-BL} & Synthetic slice & Derived from \texttt{Skill-Inject} & Controls the relative line position of the malicious instruction inside the skill file & Tests whether earlier injections are easier to hijack and easier to detect \\
\texttt{SI-CH} & Synthetic slice & Derived from \texttt{Skill-Inject} & Compares \texttt{description}-channel poisoning with direct \texttt{skill\_line} edits & Tests the core instruction-like carrier hypothesis \\
\texttt{MASB} & Real & \citep{maliciousskills2026} & Real malicious and benign agent skills collected from public repositories & Realism check and direct comparison with traditional IPI detectors \\
\texttt{MASW} & Real & \citep{maliciousagentskills2026,agentskills2026} & Empirical wild-skill corpus built from real agent ecosystems & Ecosystem-level generalization beyond controlled injected attacks \\
\texttt{BIPIA} & Ordinary IPI & \citep{liu2023prompt,liu2023formalizing} & Indirect prompt injection in data-like external carriers rather than reusable skills & Transfer test back to ordinary indirect prompt injection \\
\bottomrule
\end{tabular}
\captionof{table}{Benchmarks used in the revised experiment chapter. We intentionally separate controlled skill slices from real malicious-skill corpora and ordinary indirect prompt injection.}
\label{tab:exp_benchmarks}
\end{center}

\begin{center}
\centering
\scriptsize
\setlength{\tabcolsep}{3pt}
\begin{tabular}{p{2.6cm}p{4.2cm}p{7.6cm}}
\toprule
Family & Methods & What they represent \\
\midrule
Rule-oriented skill screening & \texttt{Skill-Inject}, \texttt{SkillScan}, \texttt{Malicious Or Not} & Surface or signature-driven detection that is strongest when malicious content appears as a visible anomaly \\
Skill-focused semantic / hybrid screening & \texttt{Agent Skills in the Wild}, \texttt{Malicious Agent Skills in the Wild}, \texttt{SkillProbe}, \texttt{MalSkills} & Detectors that reason over broader skill semantics, repository evidence, or hybrid features \\
Traditional indirect prompt-injection detectors & \texttt{PromptArmor}, \texttt{AlignSentinel}, \texttt{RENNERVATE} & Strong generic IPI defenses, including prompt-level filtering and attention-based token-level detection, that were not designed specifically for instruction-like skill carriers \\
Internal-signal ablations & \texttt{Only-Attention}, \texttt{Only-Representation} & Our component ablations for isolating the contribution of each expert \\
Ours & \texttt{RouteGuard} & Frozen-backbone attention + hidden-state detection tailored to instruction-like carriers \\
\bottomrule
\end{tabular}
\captionof{table}{Comparison systems used throughout the experiment chapter. The grouping follows the threat model discussed in the introduction rather than the order of prior papers.}
\label{tab:exp_baselines}
\end{center}

\section{Additional Motivation Evidence}
\FloatBarrier
The main text keeps only the sharpest motivation conclusions. This appendix restores the quantitative and qualitative details behind those claims, including directional routing shifts, position/channel effects, bucketed attack-success trends, and representative attack outputs.

\paragraph{A.1 Detailed boundary and directionality evidence.}
For the carrier contrast in Figure~\ref{fig:motivation_boundary}, benign \texttt{skillinject} carriers contain \texttt{1.8868} instruction-cue words per 100 tokens, compared with only \texttt{0.5786} for benign \texttt{BIPIA} external contexts. On the subtle \texttt{skillinject\_channel} regime, the lexical score gap is only \texttt{0.0156}, versus \texttt{0.0362} on \texttt{BIPIA}, and lexical attack miss rises to \texttt{0.9535}, versus \texttt{0.7917} on \texttt{BIPIA}. In the white-box contrast, \texttt{skill\_like} carriers attract more untrusted attention mass (\texttt{+0.0309 +- 0.0057}), but exhibit lower top-chunk margin (\texttt{-0.0054 +- 0.0034}) and higher entropy (\texttt{+0.9824 +- 0.0523}). For the attention-directionality analysis, malicious-target share increases by \texttt{+0.0014}, trusted-context share decreases by \texttt{-0.0013}, the malicious-to-trusted attention ratio rises by \texttt{+0.0172}, and trusted-to-target top-source switches increase by \texttt{+0.0044}. These statistics support the main-text claim that skill poisoning is not clean anomaly insertion, but directional control competition inside an instruction-like carrier.

\paragraph{A.2 Detailed behavior evidence.}
Figure~\ref{fig:appendix_attention_stats} reports that the correlation between relative injected-line position and paired delta AHS is \texttt{-0.1349 +- 0.0789}; the earliest by-line bucket (\texttt{0--20\%}) reaches \texttt{0.5753 +- 0.3191}, while the latest bucket (\texttt{80--100\%}) drops to \texttt{0.1211 +- 0.0532}. Channel effects are stronger still: \texttt{description} poisoning yields delta AHS \texttt{1.3075 +- 0.3566}, whereas direct \texttt{skill\_line} poisoning yields only \texttt{0.0253 +- 0.0909}. Under a strict output-level criterion, ASR rises across delta-AHS buckets from \texttt{0.2500} at \texttt{q1} to \texttt{1.0000} at \texttt{q5}, with intermediate values \texttt{0.4167}, \texttt{0.6667}, and \texttt{0.5833}. The trajectory is not perfectly monotonic, but the trend is clear: stronger hijacking is associated with a higher probability that the output is redirected toward the malicious objective.

\begin{center}
    \centering
    \includegraphics[width=0.88\columnwidth]{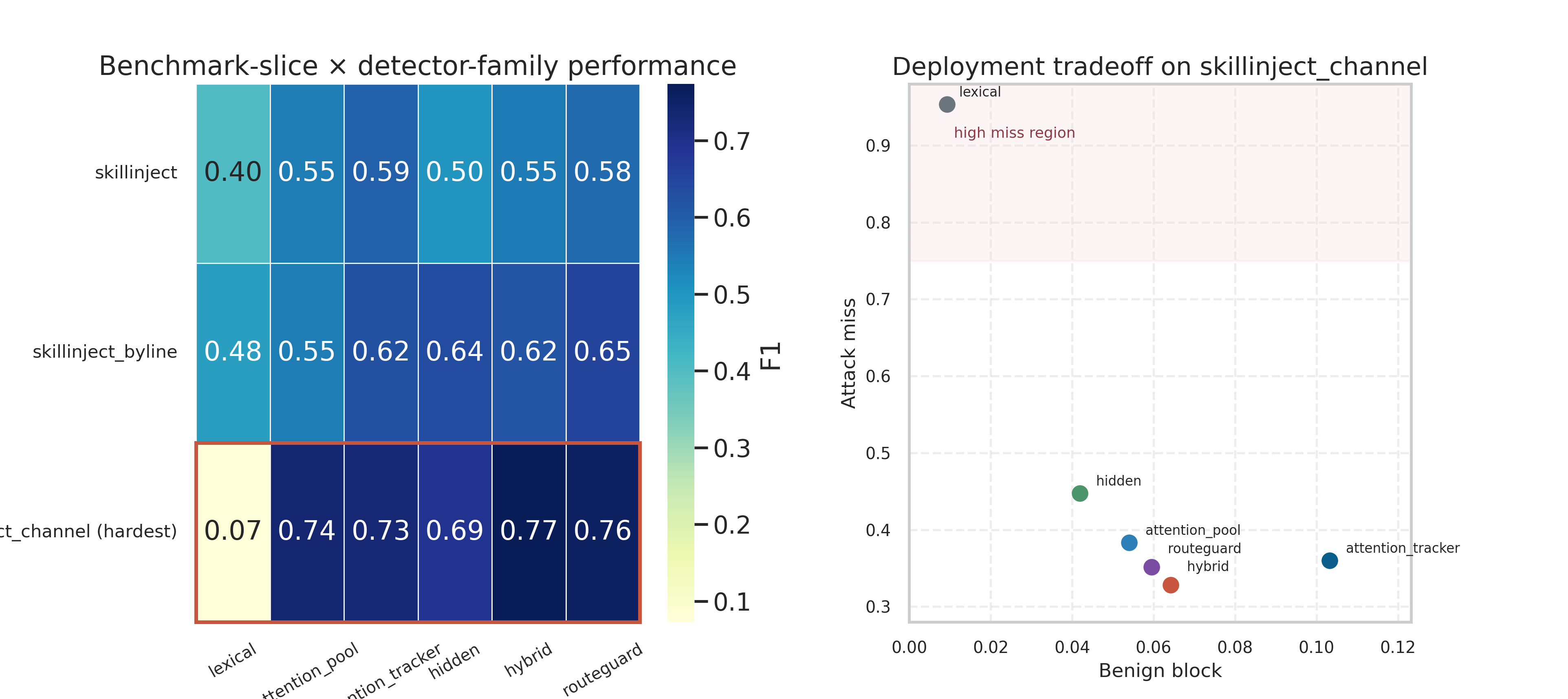}
    \captionof{figure}{Why the detector must use internal signals and fusion. Different benchmark slices privilege different signal families, and the contextual slice most relevant to skill poisoning is exactly where lexical screening collapses while internal methods remain informative.}
    \label{fig:appendix_method_motivation}
\end{center}

\begin{center}
    \centering
    \includegraphics[width=0.78\textwidth]{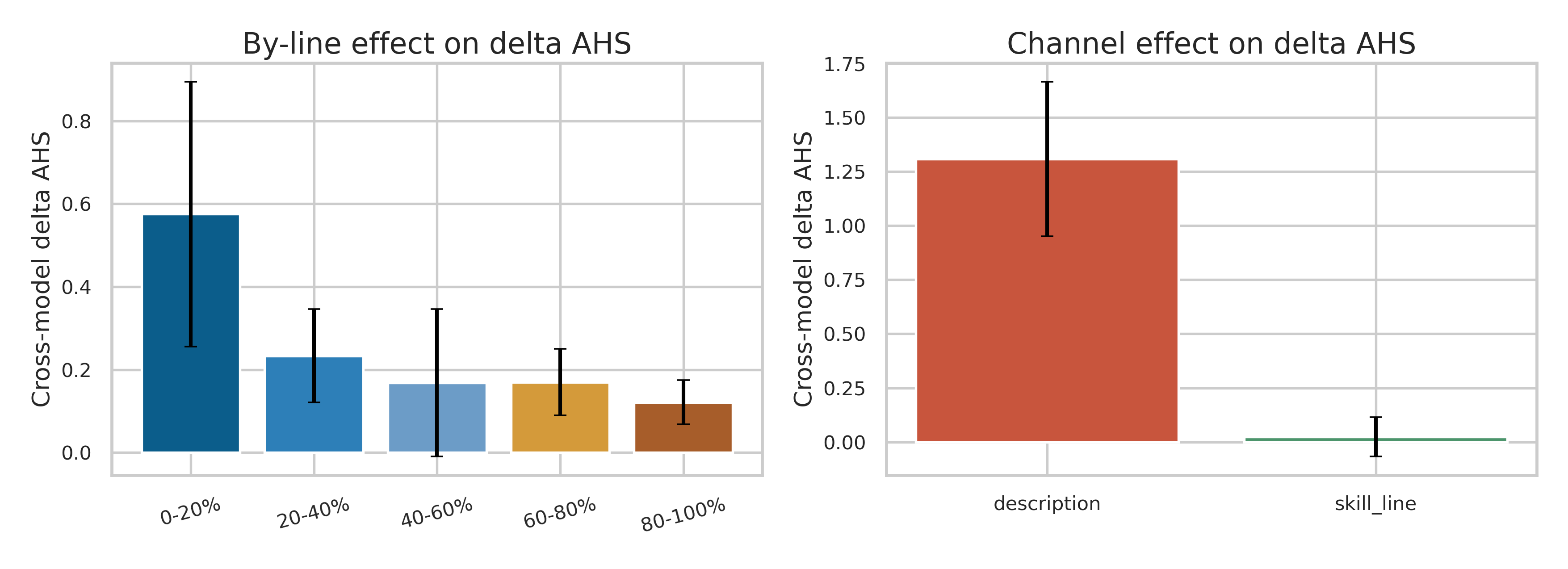}
    \captionof{figure}{Aggregated attention-hijacking statistics. Earlier injected lines induce larger paired delta AHS, contextual \texttt{description} poisoning induces much larger shifts than direct \texttt{skill\_line} edits, and the lexical-versus-internal comparison shows that the strongest internal shifts need not coincide with the strongest surface cues.}
    \label{fig:appendix_attention_stats}
\end{center}

\begin{center}
    \centering
    \includegraphics[width=0.78\textwidth]{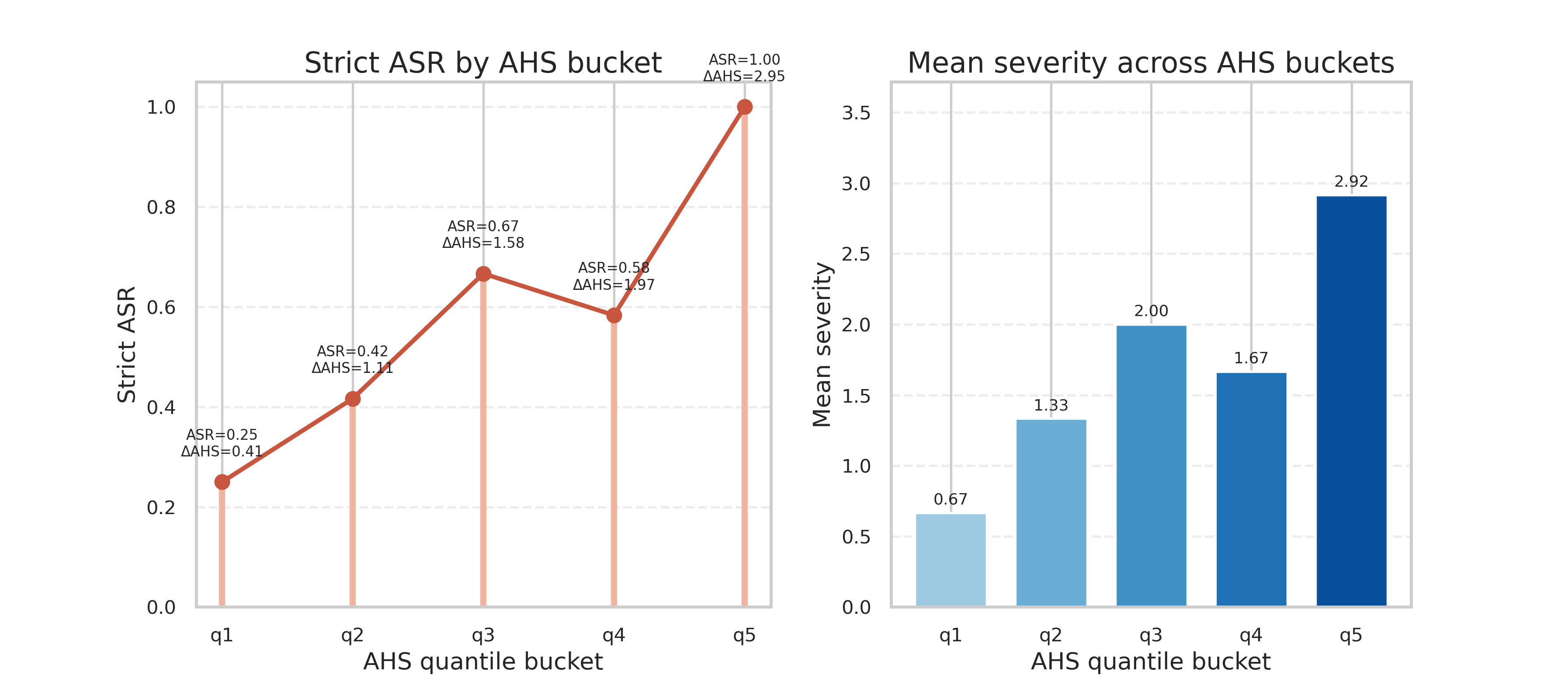}
    \captionof{figure}{Behavioral relevance of internal hijacking. ASR rises across delta-AHS buckets under a strict output-level criterion, and mean severity follows the same broad trend.}
    \label{fig:appendix_behavior}
\end{center}

\end{document}